\documentclass{sf2a-conf}
\usepackage{graphicx}
\usepackage{natbib}

\begin{document}

\TitreGlobal{SF2A 2008}
\title{Radiation processes around accreting black holes}
\author{Belmont, R.}\address{Centre d'Etude Spatiale des Rayonnements (OMP; UPS; CNRS), 9 avenue du Colonel Roche, BP44346, 31028, Toulouse Cedex 4, France}
\author{Malzac, J.$^1$}
\author{Marcowith, A.}\address{Laboratoire de Physique Th\'eorique et d'Astroparticules, IN2P3/CNRS, Universit\'e MontpellierII, CC 70, place Eug\`ene Bataillon, F-34095 Montpellier Cedex 5, France}
\runningtitle{Radiation processes in hot plasmas}
\setcounter{page}{237}
\index{Belmont, R.}
\index{Malzac, J.}
\index{Marcowith, A.}
\maketitle

\begin{abstract}
Accreting sources such as AGN, X-ray binaries or gamma-ray bursts are known to be strong, high energy emitters. The hard emission is though to originate from plasmas of thermal and/or non-thermal high energy particles. Not only does this emission allow to probe the unique properties of the matter in an extreme environment, but it also has a crucial backreaction on the energetics and the dynamics of the emitting medium itself. Understanding interactions between radiation and matter has become a key issue in the modelling of high energy sources. Although most cross sections are well known, they are quite complex and the way all processes couple non-linearly is still an open issue. 

We present a new code that solves the local, kinetic evolution equations for distributions of electrons, positrons and photons, interacting by radiation processes such as self-absorbed synchrotron and brems-strahlung radiation, Compton scattering, pair production/annihilation, and by Coulomb collisions. The code is very general and aimed to modelled various high energy sources. As an application, we study the spectral states of X-ray binaries, including thermalization by Coulomb collisions and synchrotron self-absorption. It is found that the low-hard and high-soft states can be modelled with different illumination but the same non-thermal acceleration mechanism.
\end{abstract}

\section{Introduction}

The hard emission of high energy sources such as X-ray binaries, AGN or gamma-ray bursts is expected to originate from high energy particles. In spite of large effort, understanding the properties of relativistic plasmas remains a challenge. At these high energies, Coulomb collisions become inefficient and the corresponding thermalisation timescale becomes longer than others, including radiation processes timescales. Then, kinetic and radiation processes contribute significantly to shaping the particle distribution; and understanding the highly non-linear, coupled evolution of particles and photons becomes a complicated issue that is best addressed numerically. 

Although Monte Carlo simulations allow to deal with complex geometries \citep{PSS80,Stern95,Malzac00}, they are often too time-consuming to be used for exploring the parameter space or real data fitting. For such purpose, codes solving the kinetic equations are more efficient for they can use simple prescriptions to account crudely for geometric effects \citep{LZ87,Coppi92,NM98,PW05}.

In the fist section, we present a new kinetic code developed to address the modelling of high energy plasmas.  In the second section, we investigate the thermalization by synchrotron self-absorption in the corona of accreting black-hole and we apply the results to the X-ray binary Cyg-X1. 

\section{Simulating radiation and kinetic processes in high energy sources}
\subsection{Principle}
The code developed is a one-zone code. Its abandons the detailed description of the sources geometry. Instead it focuses on the microphysics of radiation and kinetic processes and assumes a homogeneous sphere of fully magnetised plasma. The plasma properties are described by the distributions of particles and photons. Isotropy is assumed so that both particle and photon populations are described by only one-dimension energy distributions, which enables fast computation. Three distributions are considered, for photons, electrons and positrons. No assumption is made on the shape of the distributions (such as thermal or power-law for example) and the exact distributions are computed self-consistently from the microphysics. A thermal distribution of hot proton is also considered to account for Coulomb heating by hot protons. The code is time dependent and evolves simultaneously the three distributions. 

\subsection{Microphysics}
At present stage, the code accounts for Compton scattering, cyclo-synchrotron pair production and annihilation, self-absorbed radiation, self-absorbed e-p bremsstrahlung, and Coulomb collisions (e-e and e-p). The exact cross sections are used for Compton scattering and pair production/annihilation.  To achieve good accuracy in the computation of Compton scattering, a dedicated treatment is also used, which combines integration of the Klein-Nishina cross section over the distribution for large angle scattering and a Fokker-Planck treatment for small angle scattering. In addition to these elemental radiation processes, the code allows for photon injection to reproduce external illumination. It also includes prescriptions for additional heating/acceleration: the simulated plasma can be heated by thermal heating and/or non-thermal acceleration. The former process is modelled by an artificial Coulomb heating, the efficiency of which is imposed. The latter can be modelled either by injecting physically high energy particles into the system, or by second-order Fermi acceleration with a threshold. In contrast to previous studies our modelling of thermal and stochastic acceleration are fully consistent. 

\subsection{Numerical methods}
The system evolution is described by a set of 3 coupled, integro-differential equations, one for each species. Solving numerically these equations is not trivial. First, the nature of integro-differential equations is very different to that of usual differential equations and the associated numerical schemes have been much less studied. Then, the energy ranges involved in high energy sources are wide: typically 20 orders of magnitude for Blazars spectra for example. The non-local nature of radiation processes implies to combine very small and very large terms, which is numerically challenging. Last, the various timescales involved can be drastically short. Photons for example can get a large amount of energy in one single Compton scattering event. To deal with these problems, the code uses a second order in energy, semi-implicit first order in time scheme and stores the cross sections as large tables.

The use of exact cross sections or suitable combinations of asymptotic expressions and the choice of specific numerical schemes enable the use of the code over very large energy ranges for photons (from radio or lower energy bands to TeV energies or higher) and for particles (from $p/mc=10^{-7} $ to $p/mc=10^7$). This numerical improvement and the implementation of numerous microphysical processes allow to use the code in many astrophysical applications as X-ray binaries, AGN, gamma-ray bursts... The code has now been tested extensively. Tests and methods are presented in more detail in \citet{Belmont08}.

\section{The spectral states of accreting black holes}

\subsection{The Low-hard and High-soft states}
As a first application of the code, we present a study on the spectral states of galactic accreting black holes. X-ray binaries exhibit a complex variability in the light curves and the spectra. Among other spectral states, two canonical states have now been identified in many sources: the so-called low-hard (LH) and high-soft (HS) states. The LH state has usually a rather low luminosity ($L< 0.1\% L_{\rm Edd}$) and a hard spectrum (with a photon index of $\Gamma\approx 1.5-2$). It is well reproduced with a thermal Comptonization model in a corona with a temperature $k_BT\approx100$~keV.  The HS state has a higher luminosity ($L\sim 1 \% L_{\rm Edd}$). The spectrum is composed by a strong excess at a few keV and a soft high-energy tail ($\Gamma\approx2.5-3.5$) extending at least to MeV energies. It cannot be reproduced by a thermal Comptonization model. Instead, it is modelled as the sum of a disk black-body emission and its Comptonization by accelerated, non-thermal particles. 

The different nature of the Comptonization in both spectral states (thermal vs. non-thermal) is often assumed to originate from different heating processes \citep{Coppi92,PC98}. The corona in the LH state is assumed to be heated by a thermal mechanism such as Coulomb heating by hot protons \citep{NY94}, whereas the particles in the HS states are thought to be accelerated by non-thermals processes such as reconnection in an active corona \citep{Galeev79}.  While the change in disk luminosity is naturally explained by a change of the innermost radius of the accretion disk \citep{Esin97}, there is no explanation why one process should be switched on while the other is switched off during state transition.

\subsection{Thermalisation of non-thermal particles}
We have investigated the idea that particles may be accelerated only by non-thermal mechanisms and then thermalised with different efficiency depending on the spectral state. Particles can be thermalised by Coulomb collisions and synchrotron self-absorption. Although the latter effect is not as known as the former, it has been shown that exchange of synchrotron photons between particles can be very efficient in magnetised sources \citep{GS91,GHS98}. 
\begin{figure}[h]
\centering
\includegraphics[scale=0.4]{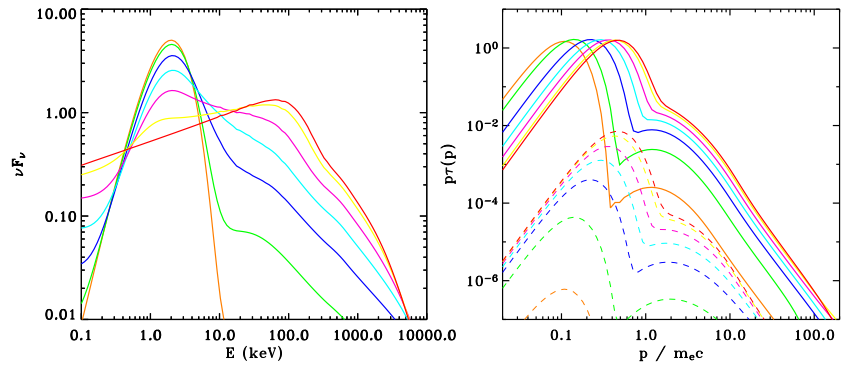}
\caption{Photon (left panel) and particle (right panel) distributions in steady state. The total energy is $L=5.5\times10^{37}$~erg~s$^{-1}$ and the fraction of energy injected as soft photons is $0, 10, 30, 50, 70, 90$, and 99\% for the red, yellow, pink, cyan, blue, green, and orange curves respectively. Dashed lines show the positron distribution. The system is characterized by its size: $R=5\times10^7$cm, its optical depths $\tau=2$, and the magnetic field strength $B=2.5\times10^6$~Gauss. The temperature of the injected soft photons is $k_B T=420$~eV and accelerated particles are injected with a power low distribution of slope $\Gamma_{\rm inj}=3$ between $\gamma_{\rm min}=1$ and $\gamma_{\rm max}=10^3$.}
\label{fig1}
\end{figure}
To study the effect of thermalisation in the different states, we ran simulations with only non-thermal acceleration.
The total injected power is the sum of the energy injected by accelerating particles and illuminating the corona. In steady state, it equals the source luminosity and was set to a constant. The faction of power injected as soft photons was varied. Figure \ref{fig1} shows steady state photon and particle distributions.
The particle distribution is composed by a high energy, non-thermal tail produced by non-thermal particle acceleration and a thermal, low energy part resulting from the thermalisation of high energy particles by synchrotron self-absorption and Coulomb collisions. For weak illumination, the thermalisation is efficient and produces a hot thermal component which produces the strong thermal comptonization spectra typical of LH states. As illumination increases, the cooling by inverse Compton increases, the temperature of the Maxwellian part decreases, and the thermal Comptonization becomes less efficient. For strong illumination, Comptonization by the non-thermal tail becomes dominant, and the spectra exhibits a power-law at high energy. As the flux of seed photons increases, the black-body component also becomes stronger, which gives eventually spectra very similar those observed in the HS state of X-ray binaries. 

\subsection{Modelling the spectrum of Cyg-X1}
\begin{figure}[h]
\centering
\includegraphics[scale=0.35]{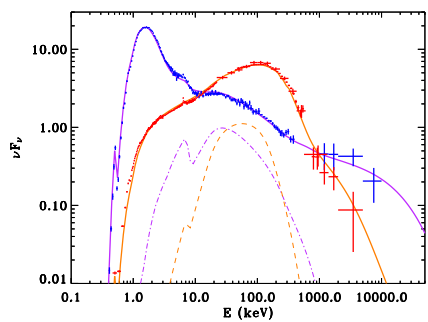}
\caption{Spectra of Cyg-X1. Crosses correspond to observations published in \citet{MacConnell02}. Solid lines are the results of two simulations with $R=10^8$cm, $B=(5.4;41)\times10^5$~Gauss, $\tau=(1.45;0.12)$, $L_{\rm phot}=(0;31)\times10^{36}$~erg~s$^{-1}$, $L_{\rm acc}=(8.8;10)\times10^{36}$~erg~s$^{-1}$, and $\Gamma_{\rm inj}=(3.5;2.1)$, for the (HS,LH) states respectively. A distance of $D=2$~kpc was assumed. Absorption with $N_{\rm H}=6\times{21}$~cm was modeld with the WABS model in XSPEC and reflection components were added to the simulations results.}
\label{fig2}
\end{figure}

We also compared these results with real data. Figure \ref{fig2} shows a comparison of two simulations with observations of Cyg-X1 in both states. 
Although these are not real fits to the data, both states are well reproduced. 

\section{Conclusion}

We presented a new code for high energy plasmas. Compared to previous codes, it describes the microphysics more accurately, more processes have been included and general numerical schemes have been implemented. This generality allows to use the code for many astrophysical situations. 

The code was used to model the spectral states of accreting black holes. We found that high energy particles can be thermalised efficiently by synchrotron self-absorption and Coulomb collisions. As a result, non-thermal particle acceleration can naturally explain even the thermal, LH state of X-ray binaries and the state transition only results from a change in the illumination by the accretion disk. Such model was successfully applied to observations of cyg-X1 and more accurate data fitting will provide strong constrains on the source parameters.





\end{document}